\documentclass[aps,a4paper,showpacs,prc,floatfix,twocolumn,amsmath,amssymb]{revtex4-1}
\usepackage{graphicx}
\usepackage{epsfig}
\usepackage{ulem}
\usepackage{morefloats}
\usepackage{rotating}
\usepackage{color,ulem}

\begin{document}

\title{Critical properties of calibrated relativistic mean-field models for the transition
  to warm, non-homogeneous nuclear matter}
\author{Olfa Boukari$^1$}
\author{Helena Pais$^2$}
\author{Sofija Anti\'c$^3$}
\author{Constan\c ca Provid\^encia$^2$}
\affiliation{$^1$ISEPBG-Soukra, University of Carthage, Avenue de la R\'epublique BP 77 -1054 Amilcar, Tunisia. \\
$^2$CFisUC, Department of Physics, University of Coimbra, 3004-516 Coimbra, Portugal. \\  
$^3$CSSM and CoEPP, School of Physical Sciences, University of Adelaide, Adelaide SA 5005, Australia. }

\begin{abstract}
 The critical properties for the transition to warm,
asymmetric,  non-homogeneous  nuclear matter  are analysed within a
thermodynamical spinodal approach for a set of well calibrated equations of state. It is
shown that even though different equations of state are constrained by the same experimental,
theoretical and observational data, and the properties of symmetric
nuclear matter are similar within the models, the properties of very
asymmetric nuclear matter, such as the one found inside of neutron stars, differ a lot for various models. Some models predict larger transition densities to homogeneous matter for
beta-equilibrated matter than for symmetric nuclear matter. Since one expects that such properties have a noticeable impact on the the evolution of either a supernova or neutron star merger, this different behavior should be understood in more detail.

\end{abstract}

\maketitle

\section{Introduction}

Core-collapse supernovae (CCSN) and neutron star (NS) mergers are two
astrophysical explosive events where matter can reach temperatures above $\approx
50$ MeV.  In CCSN matter, $\beta-$equilibrium is not immediately 
reached, and a fixed proton fraction  in the range  of $0<y_p<0.6$  is usually considered in  the simulations
\cite{Oertel}. In these very energetic events, light and heavy nuclear clusters are supposed to form, guiding the neutrino dynamics, and affecting, for example, the
cooling of the proto-neutron star \cite{cooling}, or the disk dissolution of a NS
merger \cite{fernandez2013,just2014,rosswog2015}. Hence, it is extremely
important for these clusters to be included in the equations of state (EoS)
for CCSN and NS mergers simulations, and to determine under which temperature, density and proton fraction, matter will be clusterized.

At subsaturation densities, nuclear matter goes through a liquid-gas phase
transition \cite{LG}. The border between the stable and unstable
matter is denoted by spinodal \cite{spinodal}, and it can be estimated
via dynamical or thermodynamical calculations.
In the first case, the
instabilities are determined from the fluctuations around
equilibrium. The zero-frequency one defines the spinodal
surface. In this approach, both the presence of
electrons and the Coulomb field can be  taken into account.
In the thermodynamical case, the region of instabilities is identified  by the
negative curvature of the free energy density, and the spinodal border
is defined by a zero curvature. Considering a calculation that does not include the electron
contribution and does not take into account the Coulomb interaction, 
the dynamical spinodal coincides with the
thermodynamical spinodal in the infinite wave length limit.
In the limit of small wavelengths, smaller than the nuclear force range, the
instability region  defined by the dynamical spinodal disappears. This
same small wavelength limit is obtained in a calculation of the
dynamical spinodal that includes  the electron
contribution and the Coulomb interaction.

While the dynamical spinodal may give
more realistic predictions for the crust-core phase transition in
neutron stars because it allows to take into account the finite range effects  of the
nuclear force and  the Coulomb
interaction, the thermodynamic spinodal still gives a good estimation,
as shown in Ref.~\cite{avancini2010,ducoin2008,ducoin2011}.
In particular, in Ref.~\cite{avancini2010}, the authors compared the
non-homogeneous to homogeneous matter transition density obtained
within the thermodynamical and  dynamical spinodals, and a more realistic approach,  a Thomas-Fermi (TF) calculation of non-homogeneous matter. Taking the TF calculation as reference, it was
shown that, for $\beta$-equilibrium matter, the dynamical spinodal gives
results comparable with TF, and the thermodynamical spinodal gives sligthly
larger ($\approx 10\%$) transition densities, \cite{ducoin2011}. Besides, for matter with a proton fraction equal to 0.3, as found in CCSN matter, the thermodynamical spinodal predicts transition
densities close to the ones of a TF calculation.
The liquid-gas phase transition calculated from
the spinodal decomposition has been used in experiments to study
the fragmentation of nuclear systems, in particular the time evolution
of a compound nucleus during heavy-ion collisions \cite{Chomaz}.

The liquid-gas phase transition also occurs in stellar matter, and
that explains why at subsaturation densities, one should expect
clusterized matter in core-collapse supernovae, neutron star mergers, and the inner
crust of neutron stars.
Light and heavy clusters should form at subsaturation densities, which in cold
catalysed neutron stars correspond to the inner crust region
\cite{avancini08}. In this case, spherical clusters form
  in the upper layers of the inner crust and in the bottom layers close to the crust-core
  transition, clusters with other geometries called pasta phases may
  arise due to a competition between the Coulomb interaction and the
  nuclear force \cite{ravenhall83}.

 Calculations seem to indicate that heavy
  clusters, including spherical clusters and pasta phases, may exist
  well above 1 MeV, as shown in studies that consider a  molecular dynamical
  description \cite{Sonoda2007}, a statistical
  description  \cite{Hempel10,Raduta10},  or a single
  nucleus approximation \cite{LS91,Shen98,avancini08,sumiyoshi19}.
  Moreover,  at finite temperatures, one expects that  light clusters, which may
  be understood as few-nucleon correlations, dominate at low densities.
At larger densities, but still   below saturation density, light
  and heavy clusters  coexist \cite{avancini17}. 
  The existence of clusterized matter is influenced by the
  temperature and proton fraction, and depends on the
  isovector properties of the nuclear matter model. 
  Non-homogeneous matter is expected below the critical end point of
  the nuclear liquid-gas phase transition, i.e.  for a temperature of
  the order of 14-17 MeV \cite{lourenco17}.

Light clusters at low densities can be introduced as independent degrees
of freedom within a generalized relativistic mean-field (RMF) approach
\cite{Typel,PaisPRC97,PaisPRL}. The calibration of the couplings of
these clusters has been performed by reproducing the equilibrium
constants extracted from heavy ion collisions (HIC)
\cite{qin12,indra,PaisPRL}. In particular, it has been shown that
it is important to take into account in medium effects when extracting
the equilibrium constants from the experimental measurements
\cite{PaisPRL}. Taking the parametrizations calibrated to the INDRA
measurements  \cite{PaisPRL}, we have shown that the fraction of light
clusters predicted up to the densities tested by INDRA are similar for
different models. However, different models predict different
dissolution densities \cite{custodio20}.

  In \cite{avancini06}, a general approach to study the liquid-gas
  phase transtion in asymmetric nuclear matter has been proposed
  within RMF models. The region of instability was identified by the negative curvature of the free energy. The approach allowed to analyse the so called
  distillation effect  discussed in \cite{Chomaz,spinodal}, which is
  occuring in asymmetric nuclear matter at the phase transition. In
  fact, the symmetry energy favors the formation  of a quite symmetric
  liquid phase, while the gas phase stays very neutron rich. In
  \cite{avancini06} it was discussed that the strength of the
  distillation effect is model dependent and, in particular, density
  dependent models show a weaker effect with respect to models with
  constant coupling constants. In \cite{avancini06}, however, a very
  restrict number of models was analysed.

  Applying the same approach, in Ref.~\cite{ducoin2008}, a comparison of the behavior of two types of phenomenological nuclear models, the non-relativistic Skyrme models and the RMF models, was performed. These two sets of models showed similar trends, although an anomalous behavior was obtained for one of the Skryrme models, the SIII model \cite{SIII}. In this model, the spinodal has a convex behavior at the upper spinodal border of symmetric matter, which we will refer in the discussion as $\rho_{sym}$.  As a consequence, for the  SIII model,  the transition from non-homogeneous matter to homogeneous matter occurs at a larger density for neutron rich matter than for symmetric matter, a behavior that none of the other models tested in that work showed. One of the characteristics of this model was its very small symmetry energy slope at saturation, $\approx 10$ MeV.  

The same formalism was applied more recently to analyse the effect of the density dependence of the symmetry energy on the instability region \cite{Alam}.   In this work, the thermodynamical instabilities were calculated for hot asymmetric
nuclear matter described  by different RMF models. The goal was to perform a more systematic study, and to determine the critical densities and proton fractions, in order to understand how sensitive these properties are to the density dependence  of the
symmetry energy, and  in particular,  to its slope at saturation density. In fact, presently there are strong constraints on the symmetry energy, both from the experimental side \cite{tsang12} and from {\it ab-initio} calculations for neutron matter that did not exist when the studies \cite{avancini06,ducoin2008} were performed.

Present simulations of CCSN or neutron star mergers are performed
taking into account realistic EoS, see for instance
\cite{steiner,fischer17,bauswein19}. While the EoS chosen are
generally calibrated at $T=0$ MeV, it is important to understand their
behavior under the extreme conditions attained in the above scenarios,
in order to   properly discuss the results of the simulations. It is
the main objective of the present study to calculate the thermodynamic
instabilities of several recently proposed  RMF models in order to
compare their finite temperature behavior, and, therefore, to
determine the finite temperature properties of nuclear models  that
have been calibrated at $T=0$ MeV. As discussed in \cite{avancini2010}
the determination of the thermodynamical spinodal allows for a good
estimation of the non-homogeneous nuclear matter inside a neutron star
or a CCSN while being numerically less demanding.

The following nuclear RMF models will be considered:  SFHo and SFHx \cite{steiner},
 FSU2R and  FSU2H proposed in \cite{tolos,tolos2},  TM1
 \cite{sugahara94}, and  the recently proposed TM1e
 \cite{shen20,sumiyoshi19}, DD2 \cite{Typel} and DDME2 \cite{ddme2},
 and  finally, D1 and D2  \cite{D1-D2}, closely related to DD2. The main conclusion of the present work is that while calibrated models behave in a very similar way at zero temperature and symmetric matter,
large differences were  identified for both the critical temperatures
and densities of  $\beta$-equilibrated matter in very asymmetric matter. In some models, like SFHo and SFHx, the onset of homogeneous matter in $\beta$-equilibrated matter occurs at  similar or larger densities than the ones found for symmetric nuclear matter. This will have consequences on the predictions of  CCSN or  NS merger simulations.

The structure of the paper is the following: in Section \ref{form}, the general formalism of RMF models and spinodal calculation are briefly introduced, Section \ref{res} discusses and compares the results on critical points, transition densities, and distilation effect between different models, and, finally, in Section \ref{conc}, a few conclusions are drawn.

\section{The formalism}\label{form}

A brief summary of the RMF formalism is given in the first part of the section, and we follow the notation previously used, see e.g. Ref.~\cite{Alam}, while the thermodynamical spinodal calculation and respective critical points are addressed in the second subsection.

\subsection{Field Theoretical Models with RMF Lagrangian}

In our set of RMF models, the nucleons, with mass $M$, interact with the  scalar-isoscalar meson field $\sigma$ with mass $m_\sigma$, the
vector-isoscalar meson field $\omega^{\mu}$ with mass $m_\omega$, and the vector-isovector meson field $\boldsymbol\rho^{\mu}$  with mass $m_\rho$.
The Lagrangian density is given by:
\begin{equation}
{\cal L}=\sum_{i=p,n} {\cal L}_i + {\cal L}_{\sigma} + {\cal
  L}_{\omega}  + {\cal L}_{\rho} + {\cal L}_{\sigma\omega\rho} \, ,
\end{equation}
where the nucleon Lagrangian reads
\begin{equation}
{\cal L}_i=\bar \psi_i\left[\gamma_\mu i D^{\mu}-M^*\right]\psi_i \, ,
\end{equation}
with
\begin{equation}
i D^{\mu}=i\partial^{\mu}-g_\omega \omega^{\mu}-
\frac{g_{\rho}}{2}  {\boldsymbol\tau} \cdot \mathbf{\rho}^\mu \, . 
\end{equation}
The Dirac effective mass is given by
\begin{equation}
M^*=M-g_\sigma \sigma  \, .
\end{equation}
In the above equations, $g_\sigma$, $g_\omega$ and $g_\rho$ are the meson-nucleon couplings, and $\boldsymbol \tau$ are the SU(2) isospin matrices.

The mesonic Lagrangians are:
\begin{eqnarray}
{\cal L}_\sigma&=&+\frac{1}{2}\left(\partial_{\mu}\phi\partial^{\mu}\sigma
-m_\sigma^2 \sigma^2 - \frac{1}{3}\kappa \sigma^3 -\frac{1}{12}\lambda\sigma^4\right),\nonumber\\
{\cal L}_\omega&=&-\frac{1}{4}\Omega_{\mu\nu}\Omega^{\mu\nu}+\frac{1}{2}
m_\omega^2 \omega_{\mu}\omega^{\mu} + \frac{\zeta}{4!}\zeta g_\omega^4 (\omega_{\mu}\omega^{\mu})^2, \nonumber \\
{\cal L}_\rho&=&-\frac{1}{4}\mathbf B_{\mu\nu}\cdot\mathbf B^{\mu\nu}+\frac{1}{2}
m_\rho^2 \mathbf \rho_{\mu}\cdot \mathbf \rho^{\mu}+\frac{\xi}{4!} g_\rho^4 (\mathbf\rho_{\mu}\rho^{\mu})^2, \nonumber\\
\end{eqnarray}
where
$\Omega_{\mu\nu}=\partial_{\mu}\omega_{\nu}-\partial_{\nu}\omega_{\mu} ,
\quad \mathbf{B}_{\mu\nu}=\partial_{\mu}\boldsymbol \rho_{\nu}-\partial_{\nu} \boldsymbol \rho_{\mu}
- g_\rho (\boldsymbol{\rho}_\mu \times\boldsymbol{\rho}_\nu)$, and $\kappa$, $\lambda$, $\zeta$, and $\xi$ are coupling constants.

The mesonic Lagrangian is supplemented with the following non-linear term that mixes  the $\sigma, \omega$, and $\mathbf{\rho}$ mesons \cite{steiner}:
%
%
\begin{eqnarray}
{\cal L}_{\sigma\omega\rho} &=&  g_{\rho}^2 f (\sigma,\omega_\mu\omega^\mu) 
\boldsymbol{\rho}^{\,\mu} \cdot \boldsymbol{\rho}_{\mu}\;.
\label{FLTL}
\end{eqnarray}
For the SFHo and SFHx models, $f$ is given by
\begin{equation}
f (\sigma, \omega_\mu\omega^\mu) = \sum_{i=1}^{6} a_i \sigma^i 
+ \sum_{j=1}^{3} b_j \left(\omega_{\mu} \omega^{\mu}\right)^{j}\;,
\label{eq:ffun}
\end{equation}
while for the FSU2R, FSU2H, TM1 and TM1e models, this function $f$ reduces to
\begin{equation}
f (\omega_\mu\omega^\mu) = \Lambda_v g_v^2 \omega_{\mu} \omega^{\mu}\;.
\label{eq:ffun}
\end{equation}
For these four models, the coupling constant of the non-linear term $\xi$ is absent.

For the density-dependent models, DD2, DDME2, and D1, the isoscalar couplings of the mesons $i$ to the baryons are written in the following way 
\begin{eqnarray}
g_{i}(n_B)=g_{i}(n_0)a_i\frac{1+b_i(x+d_i)^2}{1+c_i(x+d_i)^2} \, ,
\end{eqnarray}
and the isovector ones are given by
\begin{eqnarray}
g_{i}(n_B)=g_{i}(n_0)\exp{[-a_i(x-1)]} \, .
\end{eqnarray}
Here, $n_0$ is the symmetric nuclear saturation density, and $x=n_B/n_0$. For the D2 model, there are additional terms in the vector density because of the energy dependent self-energies, meaning that $n_B$ and $n_\omega$ are  no longer equal. For all density-dependent models, the coupling constants $k$, $\lambda$, $\xi$, and $\zeta$ are zero, together with the $f$ function.

%
The energy density ${\cal E} $ is given by:
\begin{eqnarray}
{\cal E^{NL}} &=&  \sum_{i=p,n} E_{i}+\frac{1}{2} m_{\sigma}^2 \sigma^2   - \frac{1}{2} m_{\omega}^2 \omega^{2}  - \frac{1}{2}m_{\rho}^2 \rho_0^{2}
  + \frac{\kappa}{6}  \sigma^3  \nonumber \\
&+& \frac{\lambda}{24}  \sigma ^4
 -\frac{\zeta}{24}(g_\omega\omega)^4-\frac{\xi}{24}(g_\rho\rho)^4
-g_\rho^2\rho^{2}f,
\label{eq:rhamz}
\end{eqnarray}
for the non-linear (NL) models, which includes several non-linear mesonic terms, and  by
\begin{equation}
{\cal E^{DD}} = \sum_{i=p,n} E_{i}+\frac{1}{2} m_{\sigma}^2 \sigma^2   - \frac{1}{2} m_{\omega}^2 \omega^{2}  - \frac{1}{2}m_{\rho}^2 \rho_0^{2}
 -\Sigma_0^R n_B\, , 
\label{eq:rhamz2}
\end{equation}
for the density-dependent (DD) models. $\Sigma_0^R$ is the rearrangement term  that appears only in the density-dependent models (see Refs. \cite{Typel,ddme2,D1-D2}), and is given by
\begin{eqnarray}
\Sigma_0^R=\frac{\partial g_{\omega}}{\partial n_B}\omega \, n_B + \frac{\partial g_{\rho}}{\partial n_B}\rho_0(\rho_p-\rho_n)/2-\frac{\partial g_{\sigma}}{\partial n_B}\sigma\rho_s \,.
\end{eqnarray}
In Eqs. (\ref{eq:rhamz}) and (\ref{eq:rhamz2}), the single-particle
energies $E_i$ are given by
\begin{eqnarray}
E_i=\frac{1}{\pi^2}\int dp\, p^2\, \epsilon_i^*
  \left(f_{i+}+f_{i-}\right) \, ,
\end{eqnarray}
the nucleon number density is
\begin{eqnarray}
\rho_i=\frac{1}{\pi^2}\int dp\, p^2 \left(f_{i+}-f_{i-}\right) \, ,
\end{eqnarray}
the scalar density is
\begin{eqnarray}
\rho_s^i=\frac{1}{\pi^2}\int dp\, p^2 \frac{M^*}{\epsilon_i^*} \left(f_{i+}+f_{i-}\right) \, ,
\end{eqnarray}
the  distribution functions  are  defined as 
\begin{eqnarray}
f_{i\pm}&=&\frac{1}{1+\exp
\left[{(\epsilon_i^*\mp \nu_i)/T}\right] } \, , 
\end{eqnarray}
with $\epsilon_i^*=\sqrt{p^2+{M^*}^2}$,
and the nucleons effective chemical potential as
\begin{equation}
\nu_i=\mu_i - g_{v} V_0 - {g_{\rho}}~  t_{3 i}~ b_0 -\Sigma_0^R,
\end{equation}
where $t_{3i}$ is the third component of the isospin operator, and the
rearrangement term is included only  for the DD models.
The entropy density $\cal{S}$ is calculated from
\begin{eqnarray}
\cal{S}&=&  -\sum_{i=n,p}\int \frac{d^3 p}{4\pi^3} ~
\left[ f_{i+}  \ln f_{i+}   +\left(1- f_{i+} \right) \ln
            \left(1-f_{i+}  \right)\right. \nonumber\\
&+&
\left.( f_{i+} \leftrightarrow f_{i-} )
\right].
\end{eqnarray}
The free energy density $\cal F$ is then obtained from the thermodynamic relation
\begin{equation}
{\cal{F}}={\cal{E}}- T\cal{S} \, .
\end{equation}


\subsection{Stability Conditions}

In the present study, we determine the region of instability of
nuclear matter constituted by protons and neutrons by calculating the
spinodal surface in the $(\rho_p, \, \rho_n,\, T)$ space. Stability
conditions for asymmetric matter  impose that the curvature matrix of
the free energy density \cite{avancini06}
\begin{equation}
{\cal C}_{ij}=\left(\frac{\partial^2{\cal F}}{\partial \rho_i\partial\rho_j}
\right)_T,
\label{stability}
\end{equation}
or, equivalently,
 \begin{equation}
\mathcal{C}=\left( 
\begin{array}{ll}
\frac{\partial \mu _{n}}{\partial \rho _{n}} & \frac{\partial \mu _{n}}{
\partial \rho _{p}} \\ 
\frac{\partial \mu _{p}}{\partial \rho _{n}} & \frac{\partial \mu _{p}}{
\partial \rho _{p}}
\end{array}
\right) ,  \label{stability1}
\end{equation}
 is positive.
The stability conditions impose
$\mbox{Tr}(\mathcal{C})>0$ and $\mbox{Det}(\mathcal{C})>0$, which is equivalent to the requirement that the two eigenvalues
\begin{equation}
\lambda _{\pm }=\frac{1}{2}\left( \mbox{Tr}(\mathcal{C})\pm \sqrt{\mbox{Tr}(
\mathcal{C})^{2}-4\mbox{Det}(\mathcal{C})}\right) 
\label{eigenv}
\end{equation}
are positive. The largest eigenvalue, $\lambda_+$, is always positive, and  the
instability region is delimited by the surface  $\lambda_-=0$. Interesting information
is given by the associated eigenvectors $\boldsymbol{\delta\rho^\pm}$, defined as
\begin{equation}
\frac{\delta \rho_p ^{\pm }}{\delta \rho _{n}^{\pm }}=\frac{\lambda^{\pm }-
\frac{\partial \mu _{n}}{\partial \rho _{n}}}{\frac{\partial \mu _{n}}{
  \partial \rho _{p}}}.
\label{drp/drn}
\end{equation}
In particular, the eigenvector associated with the eigenvalue that defines the
spinodal surface determines the instability direction, i.e. the direction
along which the free energy decreases. 

The critical points for different temperatures $T$, which are important for the definition of conditions under which the system is expected to clusterize, are also going to be calculated. These points satisfy
simultaneously  \cite{reid,avancini06}
\begin{eqnarray}
\mbox{Det}(\mathcal{C})&=&0\\
\mbox{Det}(\mathcal{M})&=&0,
\end{eqnarray}
with
\begin{equation}
\mathcal{M}=\left( 
\begin{array}{ll}
 {\cal C}_{11} & {\cal C}_{12}\\
 \frac{\partial {|\cal C}|}{\partial \rho_p}& \frac{\partial {|\cal C}|}{\partial \rho_n}
\end{array}
\right) .  
\label{critical}
\end{equation}

At a fixed  temperature, the critical  points, defined by the  pairs
  \begin{equation}
    \left(\rho_{p,c},\rho_{n,c}\right)
    \quad \mbox{ or }\quad
  \left(\rho_c, y_{p,c}\right)
\label{crit1}
\end{equation}
with
\begin{equation}
  \rho_c=\rho_{p,c}+\rho_{n,c}, \qquad y_{p,c}=\rho_{p,c}/\rho_{c},
  \label{crit2}
\end{equation}
 represent the points where the spinodal and the binodal are coincident,
  and correspond to the points of the spinodal section with
  maximum pressure (there are two pairs, which are symmetrical with respect to the
  $\rho_p=\rho_n$ line). For cold matter, the
  ($\rho_p,\rho_n$) line defined by the $\beta$-equilibrium condition
  crosses the spinodal very close to the  $T=0$ critical point, and,
  therefore, the crust-core transition density is well estimated from the crossing
of the   $\beta$-equilibrium 
  ($\rho_p,\rho_n$) line and the spinodal section.

The thermodynamical spinodals and respective  critical points are going to be
calculated for a series of the introduced RMF models in the next section.

\section{Results and discussion}\label{res}

In this Section, we start by elaborating in more detail on the models we use. For each of them, we calculate the thermodynamic instability regions, the
critical points, the transition densities, and the isospin distillation effect for a given temperature. To conclude, a discussion of
the results will be presented.

\subsection{Models}

In the present study we consider a set of RMF models calibrated to
properties of nuclei and  nuclear matter. These models fall into two
different types: one with density-dependent couplings, DD2, DDME2, D1, and D2, which we  designate by DD models, and the other with non-linear couplings, SFHo, SFHx, FSU2R, FSU2H, TM1, and TM1e, which we
designate by NL models.

In Table \ref{tab1}, some symmetric nuclear matter
properties calculated at saturation density are given for the all the models
that we explore.

\begin{table}[htb]
\caption{\label{tab1}
The symmetric nuclear matter properties at saturation density for the models under study: the nuclear saturation density $\rho_0$, the binding energy per particle $B/A$, the incompressibility $K$, the symmetry energy $E_{sym}$, the slope of the symmetry energy $L$, and the nucleon effective mass $M^{*}$. All quantities are in MeV, except for $\rho_0$ that is given in fm$^{-3}$, and the effective nucleon mass is normalized to the nucleon mass.}
\begin{ruledtabular}
\vspace{0.5cm}
\begin{tabular}{ccccccc}
 Model    &  $\rho_0$ & $B/A$ &  $K$ & $ E_{sym}$ & $L$ & $M^{*}/M$\\
\hline
SFHo  & 0.158  & 16.13 & 243  &  31.4 &  44 & 0.76  \\
SFHx  & 0.16  & 16.16  & 261  &  27 &  43 & 0.71 \\
FSU2R & 0.15  & 16.28  & 238  &  30.7  & 47 & 0.59  \\
FSU2H & 0.15  & 16.28  & 238  &  30.5  & 45 & 0.59  \\
TM1    & 0.145 & 16.3   & 281    &  36.9  & 111 & 0.63 \\
TM1e    & 0.145 & 16.3   & 281    &  31.4  & 40 & 0.63 \\
DDME2 & 0.152   & 16.14   & 251  &  32.3 & 51 & 0.57 \\
DD2   & 0.149   & 16.02   & 243  &  31.7 & 58 & 0.56 \\
D1    & 0.15    & 16.0   & 240    & 32.0  & 60 & 0.56 \\
D2    & 0.146 & 16.0   & 240    &  32.0  & 60 & 0.56 \\
\end{tabular}
\end{ruledtabular}
\end{table}

Concerning the NL models, SFHo and SFHx  \cite{steiner}  include several non-linear
terms of higher order.  They were constructed in such a way that they satisfy
constraints coming  from nuclear masses, giant monopole resonances,
and binding energies and charge radii of $^{208}$Pb and
$^{90}$Zr. Besides, they satisfy the 2-$M_\odot$ constraint
\cite{2Mstars}, and the pressure of neutron matter is always positive
and increasing. 

FSU2H and FSU2R were calibrated in order to reproduce
the properties of finite nuclei, constraints from kaon production and
collective flow in HIC, and to predict neutron matter pressures consistent
with effective chiral forces in \cite{tolos,tolos2}. Both models reproduce 2$M_\odot$
stars, have a symmetry energy and
its slope at saturation consistent with current laboratory
predictions, and their neutron skin thickness is compatible with
several experiments, both for $^{208}$Pb and for $^{48}$Ca, as from
measurements of the electric dipole polarizability of
nuclei.  The main difference between both is the softer symmetry
energy of FSU2H, with a slope at saturation 5\% smaller.

TM1e \cite{shen20,sumiyoshi19}  accurately describes finite nuclei, gives two
  solar-mass neutron stars and radii compatible with the latest
  astrophysical observations by NICER \cite{nicer}. Its 
  symmetry energy and slope at saturation 
  are also consistent with astrophysical observations
  and terrestrial nuclear experiments
  \cite{Oertel,tews17,birkhan17}, while TM1  \cite{sugahara94} fails these
  constraints, and, in particular, has a very large symmetry energy
  and  slope at  saturation.

With respect to the density-dependent models, D1 and D2  \cite{D1-D2} are close
to DD2 \cite{Typel}, which was fitted to properties of nuclei and reproduces
2$M_\odot$ stars. D2 includes an energy dependence, that was fitted to
the optical potentials \cite{optical}.  This model does not reach the
2-solar-mass constraint since the EoS becomes very soft when the
optical potential constraint is satisfied. DDME2 \cite{ddme2} was adjusted to
reproduce the properties of symmetric and asymmetric nuclear matter,
binding energies, charge radii, and neutron radii of spherical
nuclei.

\begin{figure}
\begin{tabular}{c}
\includegraphics[width=0.9\linewidth]{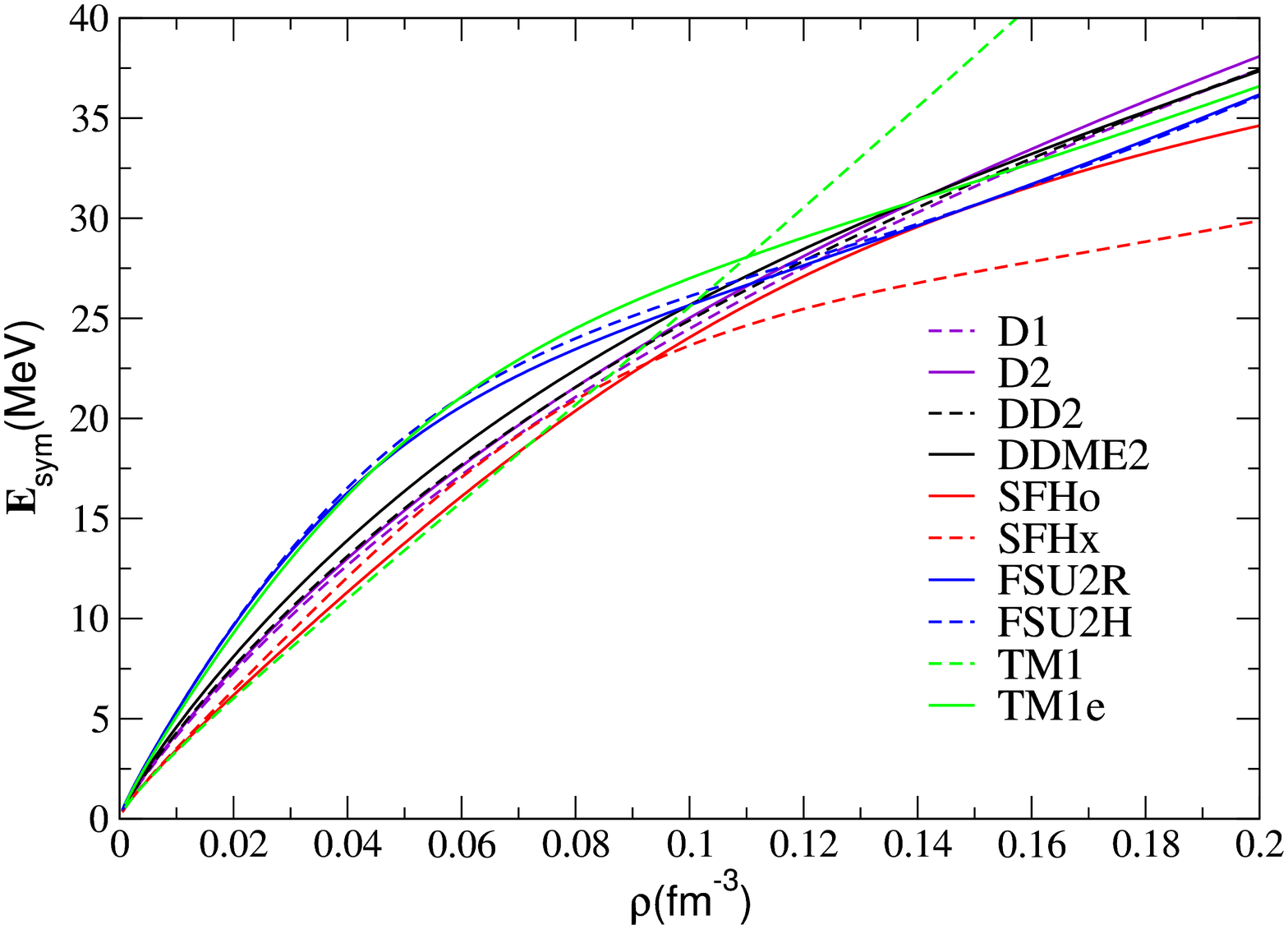}\\
\includegraphics[width=0.9\linewidth]{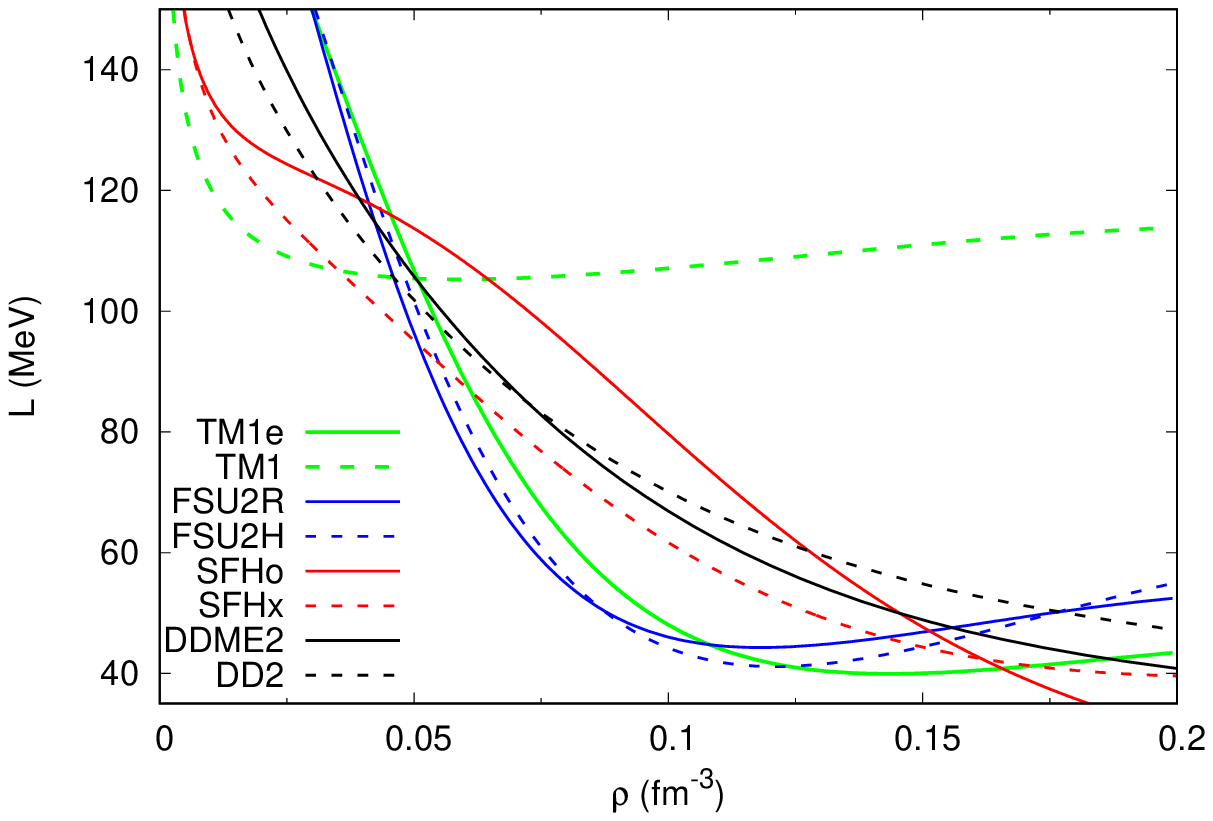}\\
\includegraphics[width=0.9\linewidth]{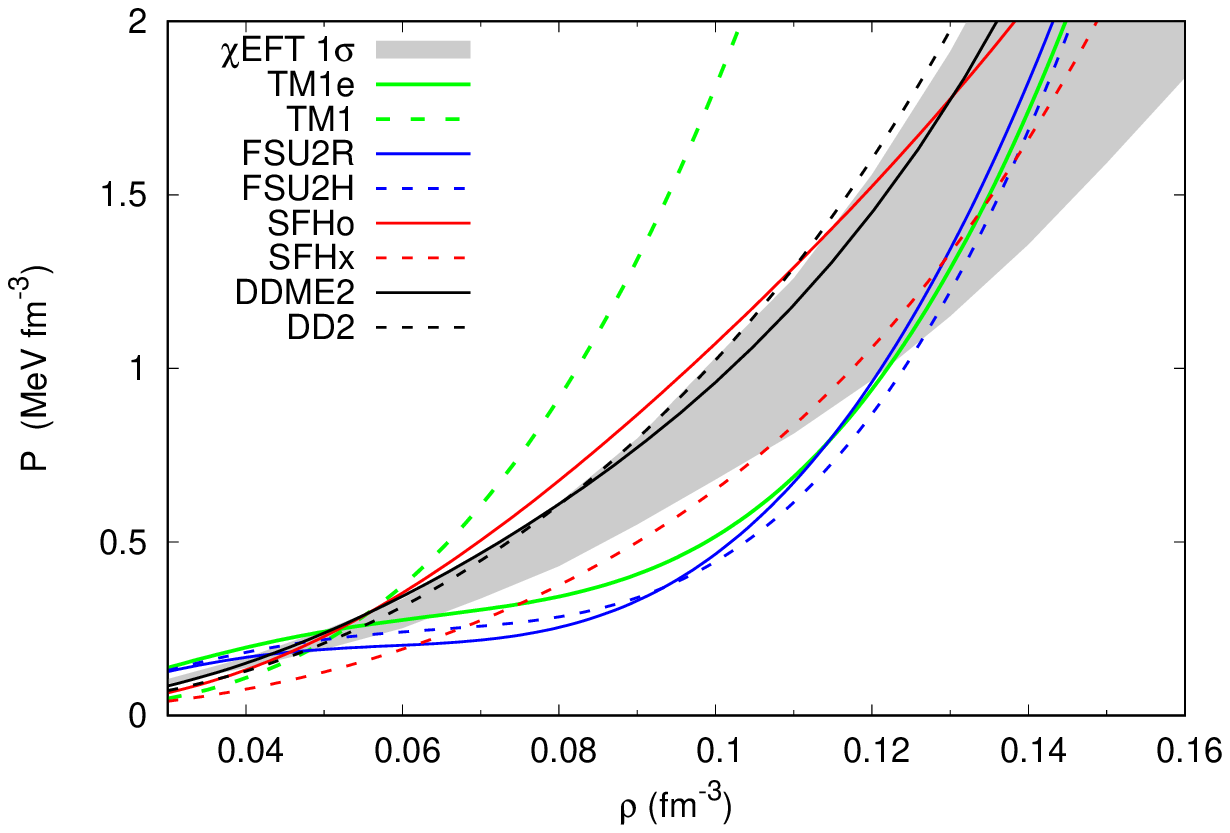}
\end{tabular} 
\caption{(Color online) The symmetry energy (top), the symmetry energy slope (middle), and the neutron matter
    pressure (bottom) as a function of the density for the   models
  under consideration. The grey band in the bottom panel represents the 1$\sigma$ constraint from chiral effective field theoretical calculations \cite{Hebeler2013}.  }
\label{fig1a}
\end{figure}

In order the better understand the isovector properties of these
models besides their properties at saturation density, in
Fig. \ref{fig1a} the
 symmetry energy  (top panel), its slope (middle panel),  and the
 neutron matter pressure (bottom panel) are plotted as a function of
 the baryonic density. In the bottom panel, we also
 include the 1$\sigma$ constraint imposed on the
 pressure of neutron matter EoS by chiral effective field theoretical
 ($\chi$EFT) calculations \cite{Hebeler2013}. Some conclusions may be drawn: SFHo is
the model that presents a softer symmetry energy above $\approx
0.5\rho_0$ and, even below this density, it is only SFHx that  is
slightly softer. While DDME2, DD2, SFHo and SFHx are quite similar
below $0.5\rho_0$,  FSU2R, FSU2H and TM1e are  clearly stiffer in this
range of densities. TM1 has an almost linear behavior with density,
presenting the smallest values below $\approx 0.1$ fm$^{-3}$, and the
largest above that value.  In fact,  above $\approx 0.1$ fm$^{-3}$,
all models have a similar behavior except TM1 that is much stiffer, and
SFHx that is quite soft.

Looking at the slope of symmetry energy, we see that TM1e, FSU2R and FSU2H follow the same trend, though TM1e has the lowest $L$ at saturation. TM1, on the other hand, is the only model that stands out, never coming below 100 MeV and having the steepest behavior for $\rho< 0.02$ fm$^{-3}$.  SFHx has a similar behavior compared to the density-dependent models, while SFHo deviates slightly from this trend, showing a steeper behavior.

It is also quite instructive to analyse the behavior of the
  neutron matter pressure. As expected TM1 completely misses the
  behavior of the  $\chi$EFT EoS.  However, the other models also
  present a quite diversified behavior. Density dependent models are
  the ones that best satisfy the $\chi$EFT constraints.  SFHo also follows  approximately the  $\chi$EFT EoS behavior. On the other hand, SFHx shows a quite low pressure in a considerable large range of densities, in particular, for $\rho\lesssim  0.06$ fm$^{-3}$. Finally, models TM1e, FSU2R and FSU2H show a too
  soft behavior of the neutron matter pressure with density below $\rho=0.08$ fm$^{-3}$, more
strongly the last two models: model FSU2R has an almost zero slope pressure at
$\rho\approx 0.05$ fm$^{-3}$. Above $\rho=0.1$ fm$^{-3}$, the pressure of these three models becomes too stiff.

We will discuss how these behaviors reflect
themselves on the instability regions.

\subsection{Spinodal sections and critical points}

\begin{figure}
\begin{tabular}{cc}
\includegraphics[width=0.78\linewidth]{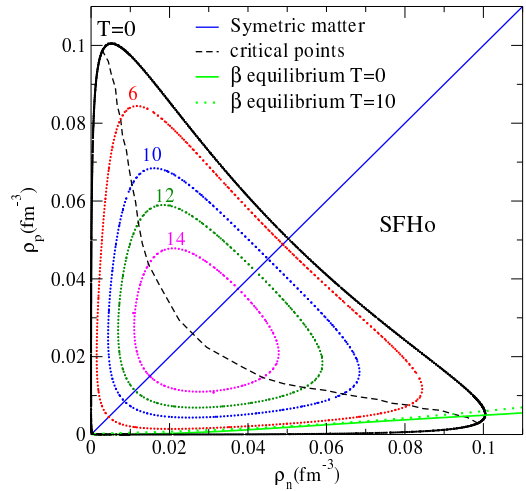}
\end{tabular}
\caption{(Color online) The spinodal regions on the $(\rho_{n},
  \rho_{p})$ plane for the SFHo model at $T = 0, 6, 10, 12$ and 14
  MeV. Also shown are the $\beta$-equilibrium EoS at $T=0$ (green solid) and
  10 MeV (green dashed), the critical points line (black dashed), and the
  symmetric matter line (blue solid).}
\label{fig1}
\end{figure}

In Fig. \ref{fig1}, we show the spinodal sections obtained with the
SFHo model at different temperatures, imposing $\lambda_-=0$, defined in Eq. (\ref{eigenv}). The larger the temperature, the
smaller the section, which will be eventually reduced to a point at the
critical temperature, that corresponds to the critical end point (CEP), and
occurs for symmetric matter. For SFHo, the CEP occurs at $T=16.14$ MeV
and $\rho=0.051$ fm$^{-3}$. It is
interesting to notice that the  $T=0$ spinodal is  convex at the
$\rho_p=\rho_n$ point.  Many of the models previously studied are
concave at this point, see for instance \cite{ducoin2008} for a
discussion. In Ref.~\cite{ducoin2008}, only the  model SIII \cite{siii} shows a
quite abnormal behavior. A consequence of this behavior is the
prediction that  highly asymmetric matter is still  non-homogeneous at
densities close, or even above, the transition density from
non-homogeneous to homogeneous matter of symmetric matter, designated in the
following as $\rho_{sym}$. However, one would expect that the
contribution of the repulsive symmetry term to the  binding energy of nuclear
matter would move the transition density to lower densities, as the
proton-neutron asymmetry increases.

In the same Figure, the EoS for $\beta$-equilibrium matter calculated
at two different temperatures, $T=0$ and 10 MeV, is also represented. The crust-core
transition density at a given temperature may be  estimated from the 
intersection of the EoS with the spinodal at that same
temperature. In Refs.~\cite{avancini2010,ducoin2011}, it was shown that this is a good
estimation although slightly larger than the values obtained within a
Thomas-Fermi or a dynamical spinodal calculations.
For the two temperatures shown, we conclude that:
i) The $T=0$ MeV EoS intercepts the $T=0$ spinodal at $ \rho_t=0.105$
fm$^{-3}$, indicating that the crust of a neutron star described by
this model extends until  approximately this density. The line
$y_p=0.5$ intercepts the spinodal at $\rho_{sym}=0.101$ fm$^{-3}$, a
density slightly smaller than $\rho_t$;
ii) the $T=10$ MeV EoS does not intercept the respective spinodal, and this indicates that
$\beta$-equilibrium matter at this temperature is homogeneous. 

The line of critical points is also displayed in the figure.  At a given temperature, these are
the two points in the spinodal section  that have maximum pressure, and where the direction of the instability is parallel to the tangent to the spinodal. This means that the pressure
above $P_{\rm max}$ belongs to the homogeneous matter phase.

\begin{figure}[t]
\begin{tabular}{c}
\includegraphics[width=0.9\linewidth]{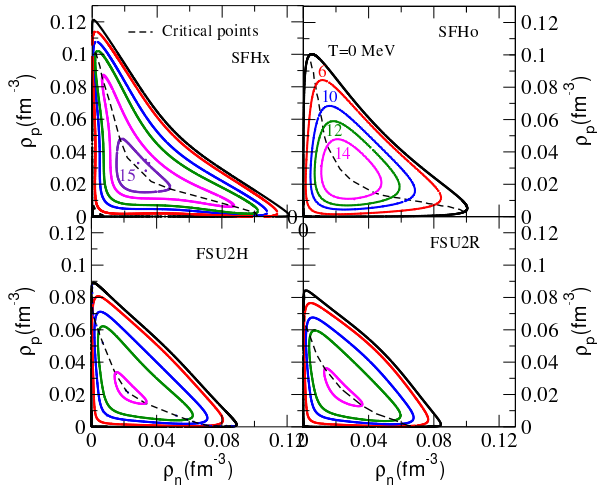}
\end{tabular}
\caption{(Color online) The spinodal sections on the $(\rho_{n}, \rho_{p})$ plane for SFHx (top left), SFHo (top right), FSU2H (bottom left) and FSU2R (bottom right) at $T = 0, 6, 10, 12$, and 14 MeV. The SFHx model is the only one that presents an unstable region at $T=15$ MeV. The critical points line is given by the black dashed line.}
\label{fig2}
\end{figure}

\begin{figure}[t]
\begin{tabular}{cc}
\vspace{2cm}
\includegraphics[width=0.6\linewidth]{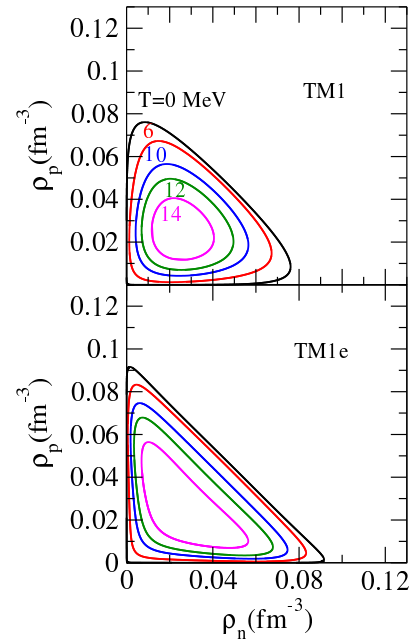}
\end{tabular}
\caption{(Color online) The spinodal sections on the
  $(\rho_{n},\rho_{p})$ plane for TM1 (top) and TM1e (bottom), at $T =
  0,\, 6,\, 10,\, 12$, and 14 MeV.}
\label{fig2a}
\end{figure}

In Table \ref{tab2}, the transition density of both
$\beta$-equilibrium matter $\rho_t$, and of symmetric matter,
$\rho_{sym}$, are given, together with the proton fraction at the
$\beta$-equilibrium transition for each model. 
For $\beta$-equilibrium matter, the transition occurs for
$y_p\ll 0.5$. All models have $\rho_{sym}>\rho_t$, except TM1e, SFHo and
SFHx, the last model having an extreme transition density of $\approx
0.12$ fm$^{-3}$. For TM1e, both densities are equal.  SFHo and
SFHx are also the models that  predict larger crust-core
transition densities.

\begin{table}
\caption{The transition density $\rho_{t}$, the correspondent proton fraction $y_{p_t}$, and the density of symmetric matter $\rho_{sym}$,  obtained at $T=0$ MeV for some of the models considered in this work. }
\begin{ruledtabular}
\begin{tabular}{cccc}
Model  &$\rho_{t}$(fm$^{-3}$) & $y_{p_t}$ & $\rho_{sym}$ (fm$^{-3}$) \\
\hline
SFHx & 0.122 & 0.041& 0.103\\
SFHo  & 0.105 & 0.047& 0.101\\
FSU2R  & 0.087& 0.045& 0.095\\
FSU2H  & 0.092& 0.046 & 0.095\\
TM1e & 0.094 & 0.050 & 0.094 \\
TM1 & 0.047 & 0.025 & 0.070 \\
DD2  & 0.081& 0.034  & 0.095\\
D1 & 0.082 & 0.032 & 0.102 \\
DDME2 & 0.087 & 0.039 & 0.099\\
\end{tabular}
\label{tab2}
\end{ruledtabular}
\end{table}

The spinodal sections obtained at
different temperatures for the NL
models we consider in this study are plotted in Figs. \ref{fig2} and \ref{fig2a}. 
SFHo and SFHx present a convex curvature at the
transition density of symmetric matter. This seems to point to some
problem in the model. They also have a bigger instability region as
compared to the other models. Comparing TM1 and TM1e, it is clearly
seen that the ones  with a smaller slope $L$ at
saturation have spinodal sections that extend to more asymmetric
matter, right up to almost the CEP, which occurs for symmetric nuclear
matter. This implies that in warm stellar matter in beta-equilibrium,
as the one found in neutron star mergers, finite clusters will appear at
larger temperatures and proton asymmetries, having direct
implications in processes like neutrino cross sections.

\begin{figure}[t]
\begin{tabular}{cc}
\includegraphics[width=0.9\linewidth]{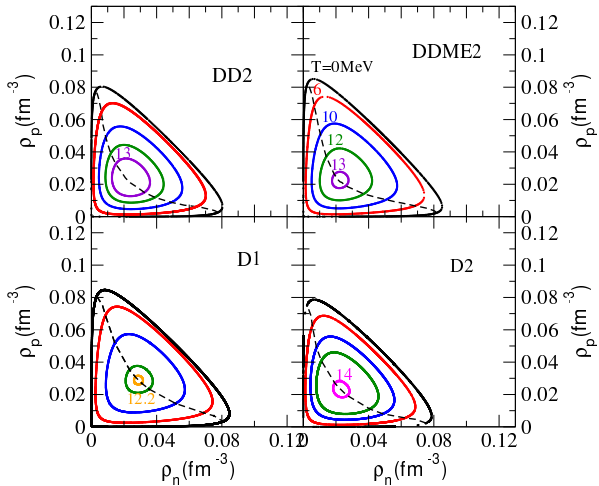}
\end{tabular}
\caption{(Color online) The spinodal sections on the $(\rho_{n},\rho_{p})$ plane for DD2 (top left), DDME2 (top right), D1 (bottom left) and D2 (bottom right) at $T = 0, 6, 10$, and 12 MeV. The smallest unstable regions shown are for $T=13$ MeV (DD2 and DDME2), 12.2 MeV (D1), and 14 MeV (D2).}
\label{fig3}
\end{figure}

On the other hand, the spinodals for DD models, which are plotted in
Fig. \ref{fig3}, show a behavior closer to the one  presented by TM1,
although having a much smaller slope $L$: the spinodal sections are
smaller, do not extend to so asymmetric nuclear matter, and they are all concave at $y_p=0.5$.

\begin{figure}
\begin{tabular}{cc}
\includegraphics[width=0.8\linewidth]{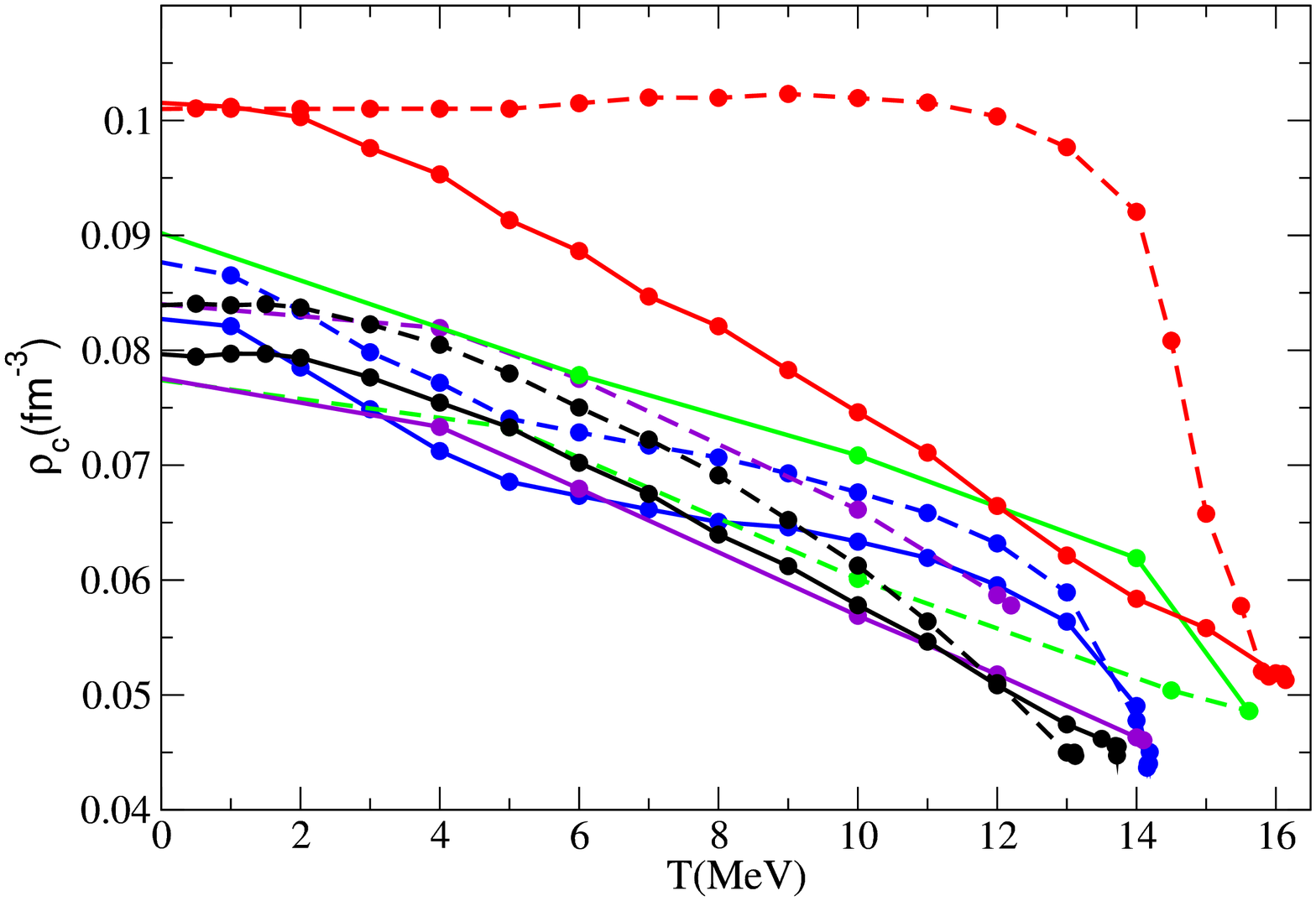} \\
$\phantom{.}$\\
$\phantom{.}$\\
\includegraphics[width=0.8\linewidth]{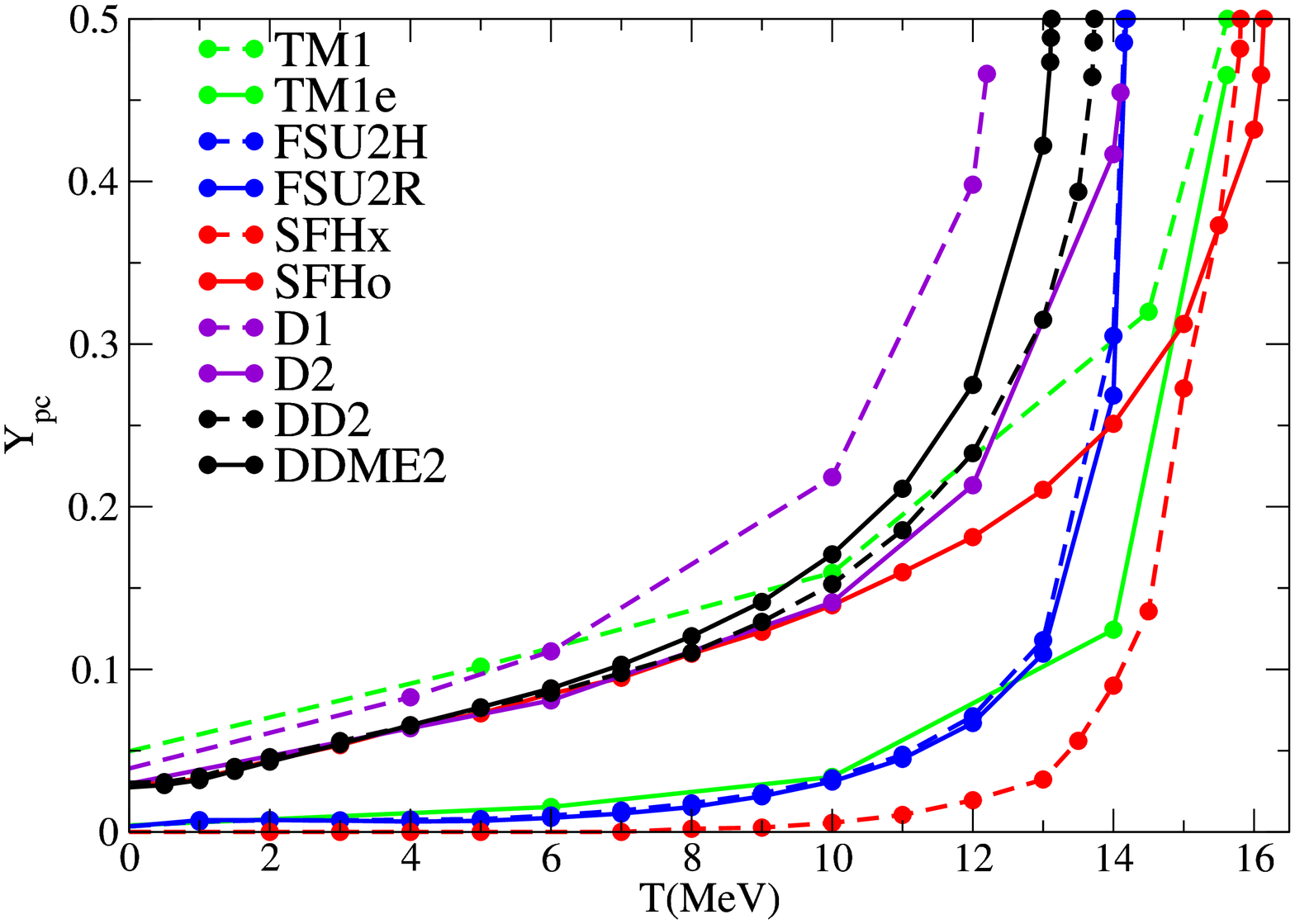}
\end{tabular} 
\caption{(Color online) The  critical density  (top) and the
  critical proton fraction (bottom), as defined in Eq. (\ref{crit2}),  as a function of the temperature $T$ for some of the models considered in this work. }
\label{fig4}
\end{figure}

The differences between the NL and DD spinodals are also clearly
seen by comparing the critical point properties at each  temperature.
In Table \ref{tab3}, we show, for several temperatures, the critical
densities and correspondent proton fractions. The same information is
given in Fig. \ref{fig4}, where the properties of the critical points
$(T, \rho_c,\, y_{pc}$) are plotted.

\begin{table}
\caption{The critical density $\rho_c$ and the correspondent proton
  fractions $y_{p_c}$ (see Eqs. (\ref{crit1},\ref{crit2}))  for different temperatures and the
  models considered in this work.}
\begin{ruledtabular}
\begin{tabular}{cccc}
    Model     &    $T$(MeV) & $\rho_{c}$(fm$^{-3}$) &  $y_{p_c}$ \\
\hline
SFHx     & 0             &0.1010    &0.0 \\
SFHo     &             &  0.1015    &0.0283\\
FSU2R    &             & 0.0827     &0.0037\\
FSU2H    &             & 0.0876     &0.0022 \\
TM1	     &             & 0.0774     &0.0496 \\
TM1e     &             & 0.0902     &0.0041 \\
D2       &             & 0.0775     &0.0296\\
D1       &             & 0.0840     &0.0390\\
DD2      &              &0.0796      &0.0302\\
DDME2    &              & 0.0839     &0.0274\\
\hline
SFHx     & 6             & 0.1015     &0.0\\
SFHo     &             &0.0886      &0.0850\\
FSU2R    &             & 0.0673     &0.0083\\
FSU2H    &              & 0.0728     & 0.0063\\
TM1e     &             & 0.0778    & 0.0154 \\
D2       &              & 0.0679     &0.0809\\
D1       &              &0.0775      &0.1110\\
DD2      &              &0.0702      &0.0855  \\
DDME2    &              & 0.0750     & 0.0882\\
\hline
SFHx     & 10            &0.1019     & 0.0056\\
SFHo     &              &0.0746   &0.1395\\
FSU2R    &             &0.0633     & 0.0304\\
FSU2H    &             & 0.0676    &0.0251 \\
TM1      &              & 0.0601  	& 0.1594 \\
TM1e     &              & 0.0708 	& 0.0339 \\
D2       &              & 0.0569    &0.1412\\
D1       &              & 0.0661    & 0.2181\\
DD2      &              &  0.0578   & 0.1523\\
DDME2    &              & 0.0612    &0.1707\\
\hline
SFHx     & 14             & 0.0920    &0.09 \\
SFHo     &             & 0.0583    &0.2509 \\
FSU2R    &             &  0.0490   & 0.2686\\
FSU2H    &             &  0.0477   &0.2607 \\
TM1e     &              & 0.0619 	& 0.1244 \\
D2       &             & 0.0463    &0.4167\\
D1       &             &- &- \\
DD2      &             & - & - \\
DDME2    &             & - &- \\
\end{tabular}
\label{tab3}
\end{ruledtabular}
\end{table}

At $T=0$ MeV, the models
SFHx, FSU2R, FSU2H and TM1e have a proton fraction at the critical point
equal to zero or very close to zero. All other models have a similar
proton fraction of the order of 0.028-0.039. At $T=6$ MeV, SFHx, FSU2R,
FSU2H, and even TM1e, still present a critical proton fraction of the
order of 0.01 or below (for SFHx it is still zero), while for all the other models, it grows up to $\approx 0.09-0.11$.

The model SFHx presents a very extreme behavior, keeping a critical proton
fraction equal to zero for $T< 10$ MeV, and a critical density of the
order of $\approx 0.1$fm$^{-3}$ for $T<12$ MeV. The models FSU2H and
FSU2R also show a critical proton fraction very close to zero for $T<8$
MeV. SFHo stands out as being the model that, after  SHFx, has the largest
critical densities, see  Fig. \ref{fig4} bottom panel.  The thermodynamic behavior of these two models
will have direct implications in the evolution of core-collapse
supernova matter or neutron star mergers since the non-homogeneous
matter will extend to larger densities and larger
temperatures. The models SFHx, FSU2R, FSU2H and TM1e predict clusterization of quite asymmetric matter for quite high temperatures. This will affect the evolution of
asymmetric stellar matter as found in neutron star mergers, or
core-collapse supernova matter after the neutrino trapped stage.

The CEP properties, i.e. the temperature and 
nuclear matter density and pressure, are given for each model in Table \ref{tab4}. At the  CEP, matter
is symmetric. The largest  CEP temperature, of the order of 16 MeV,
is obtained for SFHx and SFHo. D1 presents the smallest CEP temperature of the order of 12 MeV.

\begin{table}
\begin{ruledtabular}
\caption{The temperatures, and nuclear matter 
  density and pressure at the  CEP, for the models considered in this
  work.  At the CEP, the proton fraction is equal to 0.5.}
\begin{tabular}{cccc}
Model  &$T_c$(MeV) &$\rho_{c}$(fm$^{-3}$) &$P_{c}$ (MeV.fm$^{-3}$)\\
\hline
SFHx &15.81 & 0.052& 0.242\\
SFHo  & 16.14 & 0.051& 0.249\\
FSU2R  & 14.19& 0.045& 0.186\\
FSU2H  & 14.16& 0.044 & 0.183\\
TM1    & 15.62 & 0.049 &0.239 \\
TM1e   & 15.61 & 0.049 & 0.239 \\
DD2  & 13.73 & 0.046  & 0.178\\
DDME2 & 13.12  & 0.045 & 0.156\\
D1 & 12.22 & 0.058 & 0.187 \\
D2 & 14.14 & 0.046 & 0.193
\end{tabular}
\label{tab4}
\end{ruledtabular}
\end{table}

 In \cite{lourenco17}, the
authors made a compilation of experimental determinations of the
critical temperature of symmetric nuclear matter. The measurements
were  performed within multifragmentation reactions or fission, and the critical temperature values fluctuate between 15 and 23 MeV. However, some of the estimations are obtained
with large  uncertainties. The analysis with smaller uncertainties \cite{tc0}
 determined a critical temperature of 16.6$\pm
0.86$ MeV, considering the limiting  temperature  values
obtained in five different mass regions \cite{tc4}, where the authors obtained a temperature
above 15 MeV,  using both multifragmentation and fission processes. In Ref.~\cite{tc5}, the authors used results from six different sets
of experimental data, both involving compound nuclei or
multifragmention, and the critical temperature of 17.9$\pm 0.4$ MeV was obtained. In this last
work, the authors also determined the critical density and pressure to
be 0.06$\pm 0.01$  fm$^{-3}$, and 0.31$\pm0.07$
MeV/fm$^3$, respectively. They used Fisher's droplet model,  that was modified to account for several effects, such as Coulomb, finite size or angular momentum effects.

Regarding the models we consider in this study, 
critical temperatures above 15 MeV are obtained for TM1, TM1e, SFHo,
and SFHx. DD models have generally a critical temperature of the
order of 14 MeV, or below, and FSU2R and FSU2H have a critical
temperature just above 14 MeV. Concerning the critical density, all models
have a density $\rho_c\gtrsim 0.044$ fm$^{-3}$, but only the models
SFHx,  SFHo,  TM1, TM1e, and D1 predict a density $\gtrsim 0.05$
fm$^{-3}$, as determined in Ref.~\cite{tc5}. SFHx, SFHo, TM1 and TM1e are the models that predict a critical pressure within the range obtained in Ref.~\cite{tc5}. 

In \cite{lourenco17}, the authors have determined the CEP of several
RMF models, and, from all the models tested, only the DD models and the
models named Z271 predicted a critical temperature above 15 MeV, and the
critical pressure and density within the range proposed in \cite{tc5}. We should,
however, note that the models Z271 predict a maximum stellar mass below
1.7$M_\odot$, as shown in \cite{Pais2016}.

\subsection{Transition densities}

In the following, we discuss the transition densities from 
non-homogeneous to homogeneous matter under  different proton fraction
conditions. In the present work, we estimate the upper and lower
  density limits of the non-homogeneous region from the crossing of
  the $y_p$ line for a fixed proton fraction or the crossing of the
 $(\rho_p,\rho_n)_{\beta-\mbox{eq}}$ line with the spinodal section
  for a given temperature.

\begin{figure}
\begin{tabular}{cc}
\includegraphics[width=0.8\linewidth]{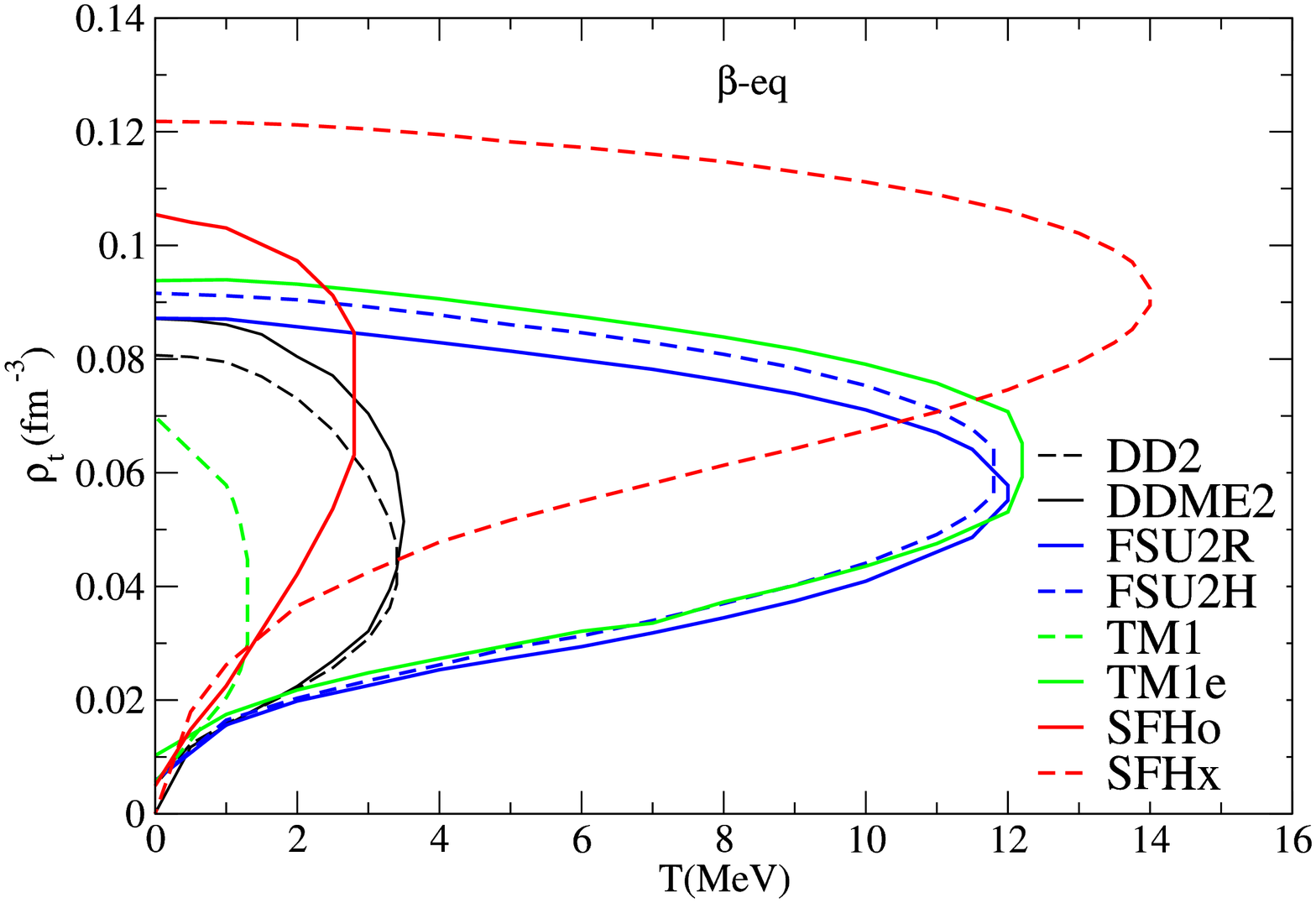} \\ 
$\phantom{.}$\\
$\phantom{.}$\\
\includegraphics[width=0.8\linewidth]{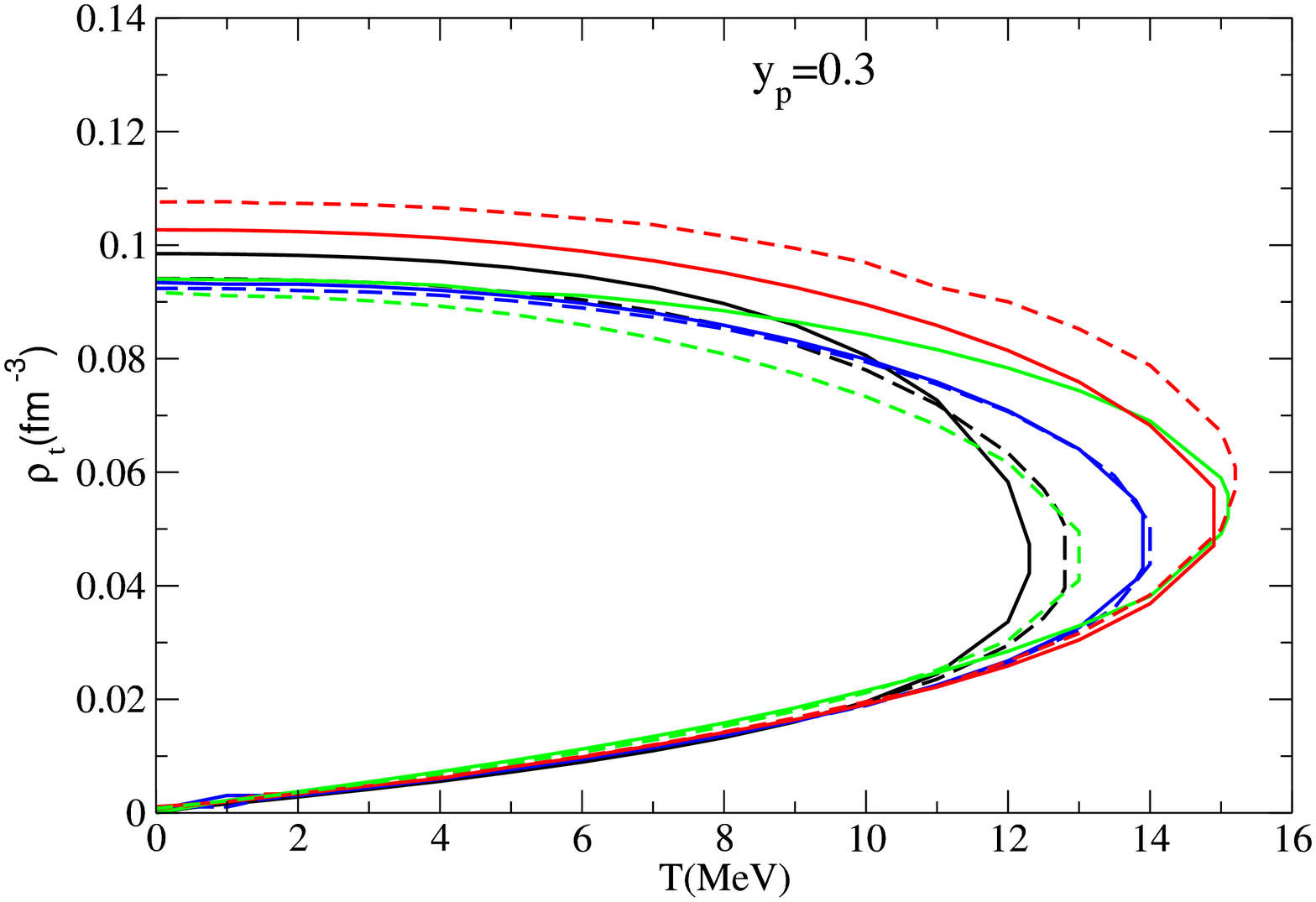}
\end{tabular} 
\caption{(Color online) The transition density, $\rho_t$, as a function of the temperature for  $\beta-$equilibrium (top) and fixed proton fraction (bottom) matter for some of the models considered in this work.}
\label{fig5}
\end{figure}

In Fig. \ref{fig5}, we show the transition densities as a function of
the temperature for two different cases: i) $\beta-$equilibrium; ii) a
fixed proton fraction of 0.3, a fraction that is representative in
core-collapse supernova matter.  Inside the represented region,
matter is, in principle, non-homogeneous. This is only an estimation of the
instability region, since we are not taking into account finite size
effects.

For  $y_p=0.3$, all models coincide at low densities and temperatures
below 10 MeV. At the upper limit, the transition densities take the
values $0.1\pm 0.01$ fm$^{-3}$ at $T=0$, and up to $T\approx 10$ MeV,
they decrease $\approx 0.02$ fm$^{-3}$. There exists experimental data that constrain matter
with this kind of asymmetry, and they show that the temperature does not affect much the
properties of nuclear matter below 10 MeV \cite{qin12,indra,pochodzalla}.
A larger  discrepancy is found for temperatures above 10 MeV. The critical temperature for this
matter asymmetry  varies between 12 and almost 16 MeV, with SFHo and
SFHx models giving the largest temperatures, and DD2 and DDME2 the
lowest ones. 

$\beta-$equilibrium matter has a much smaller proton fraction, and
there are no experimental data that can constrain the EoS of this kind of
matter. Let us, however, recall that all the models satisfy
constraints coming from chiral effective field theory calculations
for neutron matter. For $\beta-$equilibrium matter, we find that the instability region estimated by  the models considered vary a lot. SFHx predicts a $T=0$ transition density above
the one obtained for $y_p=0.3$, and a critical temperature $\approx 14$
MeV. Although with more reasonable transition densities at low
temperatures, FSU2H and FSU2R also predict very large  critical
temperatures, $\approx 12$ MeV. All the other models predict a
critical temperature of the order of 3 MeV, but show a large
dispersion on the transition density, with SFHo going above 0.1
fm$^{-3}$.
 In Ref.~\cite{sumiyoshi19}, the authors have discussed the
  influence of the density dependence of symmetry energy  on the
  supernova evolution considering the models TM1 and TM1e. They concluded that there are only minor
  effects around the core bounce and in the first milliseconds
  considering the evolution of stars with  masses of the order of 12-15
  $M_\odot$, precisely because the proton fractions are still not too far from symmetric matter at  this stage, and the predictions from both models do not differ much. However, more drastic differences between TM1 and TM1e were found at a later stage, with
TM1e giving rise to  larger neutrino emissions  and a  slower decay
of the neutrino luminosities.

As referred before, the thermodynamic calculation of the instability
regions only allows an estimation of the region where non-homogeneous
matter is expected. Finite size effects due to the finite range  of
nuclear force and  Coulomb  interaction effects will affect the
extension of the region of instability, as discussed  in
\cite{ducoin2011}. The authors showed that the transition
density obtained from a dynamical spinodal approach would predict
transition densities that are  $\approx 0.01$fm$^{-3}$ lower and
proton fractions 10\% smaller, which are good lower limit estimations, as compared to a thermodynamical spinodal calculation. A Thomas-Fermi calculation of the non-homogeneous matter may give slightly larger transition densities, as shown in \cite{avancini2010}.

\begin{figure}
\begin{tabular}{cc}
\includegraphics[width=0.9\linewidth]{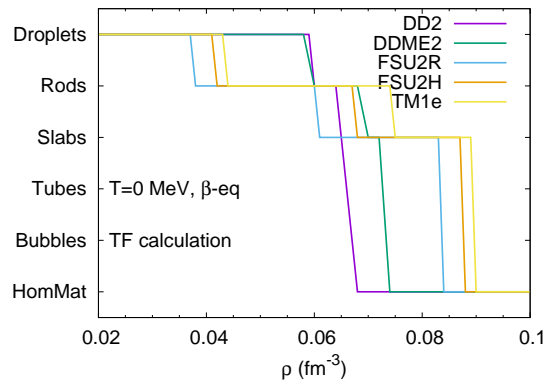}
\end{tabular} 
\caption{(Color online) The different pasta structures from a Thomas-Fermi calculation for cold $\beta-$equilibrium matter for some of the models under consideration.}
\label{fig6}
\end{figure}

In Fig. \ref{fig6}, we show the transition densities between the different
nuclear pasta phases, together with the transition density to homogeneous matter,
for five of the models under consideration. These densities were
calculated from a Thomas-Fermi approximation at $T=0$ MeV and
$\beta-$equilibrium matter \cite{grill}.  As expected, the crust-core
transitions obtained in these calculations are  lower than the ones
estimated from our thermodynamical approach, by not more than 0.01
fm$^{-3}$. It is interesting to notice that while DD2 and DDME2
predict  a large extension of the spherical clusters in the inner
crust, a shorter extension of the rod phase, and no slab phase, or a
very narrow one, the  models FSU2H, FSU2R, and TM1e predict similar extensions
of the droplet-like, rod-like and slab-like pasta structures. These
different geometries will certainly  affect the transport properties
of the neutron star inner crust.

\subsection{Distillation effect}

Transport properties are also affected by the proton content of the
gas phase, when matter clusterizes. In the following, we determine
how the system separates into two phases and the
isospin content of each. This will be achieved comparing the direction of the eigenvector of
the free energy curvature matrix associated with the negative
eigenvalue as defined in Eq. (\ref{eigenv},\ref{drp/drn})  with the direction defined by
$y_p=\rho_p/(\rho_n+\rho_p)$, see \cite{margueron03}.
If the directions are coincident, the instability   does   not change the proton fraction, and the
fluctuations that drive the phase transition are purely density
fluctuations. In \cite{margueron03}, it was shown
that  the  eigenvector associated with the instability tends to point
in the direction of increasing symmetry of the liquid phase, and,
therefore, increasing asymmetry of the gas phase.

We designate by  isospin distillation effect
the tendency of matter to separate into a low-density phase, the gas phase, that is more
neutron rich, i.e. with low proton fraction, and a high-density phase, the
clusters, with a proton fraction closer to the one of symmetric matter, i.e. with high
proton fraction. A simple way of  identifying the distillation
  effect is by the comparison of the ratio
  ${\delta \rho_p^{-}}/{\delta \rho _{n}^{-}}$ with  ${\rho_p
    }/{\rho _{n}}$. This will be used in the following to compare the
    distillation effect within the models we are discussing.

\begin{figure}
\begin{tabular}{c}
\includegraphics[width=1.0\linewidth]{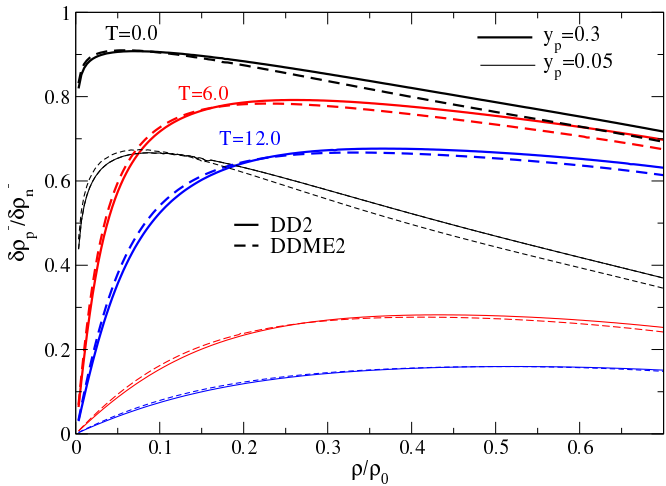}
\end{tabular} 
\caption{(Color online) The fluctuations $\delta \rho _{p}^{-}/\delta \rho _{n}^{-}$ at $T=0, 6$, and 12 MeV as function of the density,  with $y_{p} = 0.3$ (thick lines) and  $0.05$ (thin lines), corresponding to $\rho_p/\rho_n=0.43$ and 0.05, for DD2 and DDME2. }
\label{fig7}
\end{figure}

\begin{figure}
\begin{tabular}{c}
\includegraphics[width=1.0\linewidth]{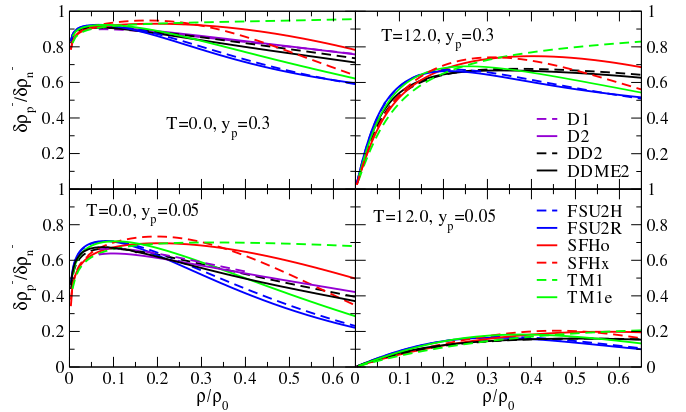}
\end{tabular} 
\caption{(Color online) The fluctuations $\delta \rho_{p}^{-}/\delta
  \rho _{n}^{-}$ at $T=0$  (left panels)  and 12 MeV (right panels)
  as a function of the density,  with $y_{p} = 0.3$ (top panels) and
  $0.05$ (bottom panels), corresponding to $\rho_p/\rho_n=0.43$ and 0.05, for FSU2R and FSU2H, SFHo and
  SFHx,  DD2 and DDME2, TM1  and TM1e. }
\label{fig9}
\end{figure}

In Fig.~\ref{fig7},  we show the isospin
distillation effect for DD2 and DDME2, by plotting the ratio of the proton to the neutron density
fluctuations associated with the negative eigenvalue inside the
instability region as a function of density, for three temperatures ($T=$0, 6 and 12 MeV) and two proton fractions, $y_p=0.05$ and 0.3, corresponding, respectively, to $\rho_p/\rho_n=0.05$ and 0.43. The proton fractions considered are representative of, respectively,
cold catalized matter in the inner crust \cite{Grill2012}, and warm matter in
protoneutron matter with trapped neutrinos just after the supernova
explosion \cite{Prakash97}.  In Fig. \ref{fig9},  the ratio of the
proton to the neutron density is plotted for all the models, for
temperatures $T=0$ and 12 MeV, and  the same proton fractions, 0.05 and 0.3.
  
The higher the ratios, the
higher the distillation effect, because  the liquid phase becomes
proton richer.  This effect  decreases when the temperature
increases, and,  except for TM1, it attains a maximum for $ 0.1\lesssim
\rho/\rho_0 \lesssim 0.3$. 
The largest differences between models are identified for the proton fraction
  0.05 and zero temperature. DD2, DDME2, D1 and D2 all show a very
  similar behavior in the whole range of densities.  In fact, all models have a similar behavior for $\rho/\rho_0 \lesssim 0.2$, and the differences occur above this
density. In particular, the ratio of the proton to the neutron density  fluctuations for
FSU2H and FSU2R (SFHo and SFHx) decreases faster  (slower) with the
density. For densities around half the saturation density, FSU2H and FSU2R have
the smallest  distillation effect. These different  behaviors reflect
the density dependence of  the symmetry energy of the models as shown
in the top and middle panels of Fig.~\ref{fig1a}: between $\rho=0.06$ and 0.1 fm$^{-3}$, these two models, together with TM1e, have the largest symmetry energy and the smallest symmetry energy slope.

\section{Conclusions}\label{conc}

In the present work, we have studied the
instability region of warm and  asymmetric nuclear matter,
considering several recently-proposed calibrated RMF models. At $T=0$ MeV,
these models have been constrained by nuclear properties, ab-initio
theoretical calculations for neutron matter, and neutron star
observations. No constaint was  imposed at finite temperature.
The thermodynamical spinodal sections in the ($\rho_p, \, \rho_n$) plane for several temperatures and the critical points have been calculated.

The main conclusions are: i) for symmetric nuclear matter, the
transition density to homogeneous matter spreads over a range
narrower than 0.01 fm$^{-3}$, $0.094<\rho_{sym}<0.103$ fm$^{-3}$;  ii) for asymmetric matter, in
particular,  for $y_p=0.3$,  the transition density to homogeneous matter
obtained from the models considered is compatible within $\approx
0.02$ fm$^{-3}$, for temperatures below 8 MeV; iii) above $T=8$ MeV, the models
differ much more, and the critical temperatures vary  in a range of 4
MeV, $12.2<T_c<16.2$ MeV; iv) properties predicted for very asymmetric matter,
as  $\beta$-equilibrated stellar matter, differ a
lot, both on the transition density, and  on the critical
temperature above which $\beta$-equilibrated matter  is not
clusterized. SFHo, SFHx, FSU2H and TM1e models predict transition densities from
clusterized matter to homogeneous matter for  $\beta$-equilibrated
matter equal or similar to the one for symmetric matter.  Since the symmetry energy contribution is a
repulsive contribution, one could expect that the  extension of the
instability region of asymmetric matter would be smaller than the one of
symmetric nuclear matter, as it happens  with all density dependent
models we have considered. The consequences of this behavior for the
evolution of neutrons stars should be understood.
It is also interesting to compare the critical temperature  of
$\beta$-equilibrated matter: models SFHx, TM1e, FSU2R and FSU2H
predict a temperature that is just $\lesssim 2$ MeV smaller than the
one obtained for symmetric nuclear matter, while all the other models
predict  temperatures  between 8 to 10 MeV smaller. Again, it is expectable that
these properties will have noticeable impact on the
 the evolution of either a supernova or neutron star
 mergers. 

 Sumiyoshi {\it et al.} \cite{sumiyoshi19} have shown, by using two
models, TM1 and TM1e \cite{shen20}, which only differ in the isospin channel, that a softer symmetry energy is responsible for
a more drastic evolution of  the  proto-neutron star with larger
neutrino emissions, giving rise to  higher neutrino luminosities and
average energies. 
Also, very recently, the  SFHo EoS has been used in several simulations of
neutron star mergers,  black hole - neutron star (BH-NS) mergers and
core-collapse supernova \cite{barbieri2019,barbieri2020,Schneider2019,Miller2019}. 
In particular, in \cite{barbieri2020}, the authors have discussed the
possibility  of  a kilonova production during  a BH-NS merger, and SFHo,
one of the preferred models, predicted smaller masses outside the blackhole.

\section*{ACKNOWLEDGMENTS}

This work was partly supported by the FCT (Portugal) Projects No. UID/FIS/04564/2019,  UID/FIS/04564/2020,  and POCI-01-0145-FEDER-029912, and by PHAROS COST Action CA16214. H.P. acknowledges the grant CEECIND/03092/2017 (FCT, Portugal). S.A. acknowledges the HGS-HIRe Abroad grant from Helmholtz Graduate School for Hadron and Ion Research.

\end{document}